\documentclass[preprint,showpacs,preprintnumbers,amsmath,amssymb]{revtex4}

\usepackage{graphicx}
\usepackage{dcolumn}
\usepackage{bm}

\renewcommand{\bar}[1]{\overline{#1}}

\begin{document}

\title{Pion-photon and photon-pion transition form factors in light-cone
formalism}
\author{Bo-Wen Xiao}
\affiliation{Department of Physics, Peking University, Beijing 100871, China}
\author{Bo-Qiang Ma}
\email{mabq@phy.pku.edu.cn}
\altaffiliation{corresponding author.}
\affiliation{ CCAST (World Laboratory), P.O.~Box 8730, Beijing 100080, China\\
Department of Physics, Peking University, Beijing 100871, China}

\begin{abstract}
We derive the minimal Fock-state expansions of the pion and the
photon wave functions in light-cone formalism, then we calculate
the pion-photon and the photon-pion transition form factors of
$\gamma ^{\ast }\pi ^{0}\to \gamma $ and $\gamma ^{\ast }\gamma
\to \pi ^{0}$ processes by employing these quark-antiquark wave
functions of the pion and the photon. We find that our calculation
for the $\gamma ^{\ast }\gamma \to \pi ^{0}$ transition form
factor agrees with the experimental data at low and moderately
high energy scale. Moreover, the physical differences and inherent
connections between the transition form factors of $\gamma ^{\ast
}\pi ^{0}\to \gamma $ and $ \gamma ^{\ast }\gamma \to \pi ^{0}$
have been illustrated, which indicate that these two physical
processes are intrinsically related. In addition, we also discuss
the $\pi ^{0}\to \gamma \gamma $ form factor and the decay width $
\mathit{\Gamma }(\pi \to \gamma \gamma )$ at $Q^{2}=0$.
\end{abstract}

\pacs{14.40.Aq; 12.39.Ki; 13.40.Gp; 13.60.Le}


\preprint{ Published in Phys. Rev. D68 (2003) 034020}




\vfill



\vfill


\vfill



\vfill 



\maketitle

\section{INTRODUCTION}

The light-cone formalism provides a convenient framework for the
relativistic description of hadrons in terms of quark and gluon
degrees of freedom \cite{Bro89}. In this formalism, the hadronic
wave function which describes a hadronic composite state at a
particular light-cone time $ \tau $ is expressed in terms of a
series of light-cone wave functions in Fock-state basis, for
example, for the pion
\begin{equation}
\left\vert \pi \right\rangle =\sum \left\vert q\overline{q}\right\rangle
\mathit{\psi }_{q\overline{q}}+\sum \left\vert q\overline{q}g\right\rangle
\mathit{\psi }_{q\overline{q}g}+\cdots,  \label{Pion Fock}
\end{equation}%
and the temporal evolution of the state is generated by the
light-cone Hamiltonian $H_{LC}^{QCD}$. Similar with $\pi ^{0}$, we
assume that the photon may also have this kind of Fock-state
expansion
\begin{equation}
\left\vert \gamma \right\rangle =\sum \left\vert q\overline{q}\right\rangle
\mathit{\psi }_{q\overline{q}},  \label{Photon Fock}
\end{equation}%
for application to QCD involved processes. To simplify the
problem, we only take into account the minimal Fock-states of the
pion and the photon, which are the lowest valence states of their
light-cone wave functions and the first order contributions in the
calculation. This consideration will be proved reasonable by the
good agreement between the numerical results and the experimental
data.

We will derive the quark-antiquark wave functions of the pion and
the photon by employing the light-cone Fock expansion of the
minimal Fock-states. Brodsky \textit{et al.} \cite{BD80,Bro2001}
have discussed the relativistic QED composite systems by giving
explicit the light-cone wave functions for the two-particle
Fock-states of the electron in QED. Along with the similar idea,
we try to obtain the Fock expansions of the photon and the pion
wave functions by calculating their vertexes in this paper. After
that, we discuss the pion-photon $(\gamma ^{\ast }\pi ^{0}\to
\gamma )$ and the photon-pion $(\gamma ^{\ast }\gamma \to \pi
^{0})$ transition form factors in the light-cone quark model.
There are many other different approaches to discuss these two
transition form factors, such as the perturbative QCD formalism by
Brodsky-Lepage~\cite{Lep80} and Cao-Huang-Ma~\cite{Cao96}, as well
as a light-front quark model by Kroll-Raulfs~\cite{Kro96} for the
$\gamma ^{\ast }\pi ^{0}\to \gamma $ transition, the QCD sum rule
calculation by Radyushkin-Ruskov~\cite{Rad97}, and a light-front
quark model by Hwang~\cite{Hwang2001} for the $\ \gamma ^{\ast
}\gamma \to \pi ^{0}$ transition form factor. However, few people
notice that these two transition form factors are physically
different although they have inherent connections. In this work,
we also try to elucidate the intrinsic relations and differences
between these two cases in the light-cone formalism, thus arrive
at a more clear understanding on the transition form factors of
$\gamma ^{\ast }\pi ^{0}\to \gamma $ and $\gamma ^{\ast }\gamma
\to \pi ^{0}$ processes.

The paper is organized as follows. In Sec.~II and Sec.~III, we
will derive the pion and the photon wave functions with helicity
structure by the relativistic vertex method of calculating the
light-cone matrix elements that developed in
Refs.~\cite{BD80,Bro2001}. In Sec.~IV, we obtain the pion-photon
and the photon-pion transition form factors by using the
pion-quark-antiquark and the photon-quark-antiquark wave
functions. In Sec.~V, we present a brief summary.

\section{THE TWO PARTICLE FOCK-STATE OF THE PION VERTEX}

We now derive the minimal Fock-state wave function of the pion in
light-cone formalism. From the Fock-state basis of the pion vertex
Eq.~(\ref{Pion Fock}), we will employ two different methods to get
the quark-antiquark wave function of the pion in this part of the
paper. One is from the light-cone quark model by taking into
account the Melosh-Wigner rotation effect
\cite{Ma93,Huang94,Xiao2002}, and another is from a full
relativistic field theory treatment of the interaction vertex
\cite{BD80,Bro2001}. It will be shown that the results from these
two methods are essentially identical.

In the light-cone quark model, the light-cone wave function of a
composite system can be obtained by transforming the ordinary
equal-time (instant-form) wave function in the rest frame into
that in the light-front dynamics, by taking into account the
relativistic effects such as the Melosh-Wigner rotation
\cite{Ma93,Huang94,Xiao2002,Ter76,Kar80,Chu88}. In the pion rest
frame $(q_{1}+q_{2}=0)$, the ordinary SU(6) quark model spin space
wave function of the pion is
\begin{equation}
\chi _{T}=(\chi _{1}^{\uparrow }\chi _{2}^{\downarrow }-\chi _{2}^{\uparrow
}\chi _{1}^{\downarrow })/\sqrt{2},  \label{instant-form}
\end{equation}%
in which $\chi _{i}^{\uparrow ,\downarrow }$ is the two-component
Pauli spinor and the two quarks have 4-momentum $q_{1}^{\mu
}=(q_{1}^{0}, \mathbf{q})$ and $q_{2}^{\mu
}=(q_{2}^{0},-\mathbf{q})$, with $
q_{i}^{0}=(m_{i}^{2}+\mathbf{q}^{2})^{1/2}$, respectively. The
instant-form spin states $|J,s\rangle _{T}$ and the light-front
spin states $ |J,\lambda \rangle _{F}$ are related by the
Melosh-Wigner rotation $U^{J}$ \cite{Wig39,Mel74}
\begin{equation}
|J,\lambda \rangle _{F}=\sum_{s}U_{s\lambda }^{J}|J,s\rangle _{T}.
\label{Wigner}
\end{equation}%
Applying the transformation Eq.~(\ref{Wigner}) on both sides of
Eq.~(\ref {instant-form}), we can obtain the spin space wave
function of the pion in the infinite-momentum frame. For the left
side, i.e., the pion, the transformation is simple since the
Melosh-Wigner rotation is unity. For the right side, i.e., the two
spin-1/2 partons, the instance-form and the light-front spin
states are related by the Melosh-Wigner transformation \cite
{Wig39,Mel74,Ma91},
\begin{equation}
\left\{
\begin{array}{lll}
\chi _{i}^{\uparrow }(T) & = & \omega _{i}[(q_{i}^{+}+m_{i})\chi
_{i}^{\uparrow }(F)-q_{i}^{R}\chi _{i}^{\downarrow }(F)], \\
\chi _{i}^{\downarrow }(T) & = & \omega _{i}[(q_{i}^{+}+m_{i})\chi
_{i}^{\downarrow }(F)+q_{i}^{L}\chi _{i}^{\uparrow }(F)],
\end{array}
\right.  \label{Melosh}
\end{equation}
where $\omega _{i}=[2q_{i}^{+}(q_{i}^{0}+m_{i})]^{-1/2}$, $
q_{i}^{R,L}=q_{i}^{1}\pm {\mathrm i} ~q_{i}^{2}$, and
$q_{i}^{+}=q_{i}^{0}+q_{i}^{3}$. Then we get the light-cone spin
wave function of the pion ($m_{1}=m_{2}=m)$ ,
\begin{equation}
\chi ^{K}(x,\mathbf{k}_{\perp })=\sum_{\lambda _{1},\lambda
_{2}}C_{0}^{F}(x, \mathbf{k}_{\perp },\lambda _{1},\lambda
_{2})\chi _{1}^{\lambda _{1}}(F)\chi _{2}^{\lambda _{2}}(F),
\label{spin}
\end{equation}%
where the component coefficients $C_{J=0}^{F}(x,\mathbf{k}_{\perp
},\lambda _{1},\lambda _{2})$, when expressed in terms of the
instant-form momentum $ q_{\mu }=(q^{0},\mathbf{q}=\mathbf{k})$,
have the forms:
\begin{equation}
\left\{
\begin{array}{lll}
C_{0}^{F}(x,\mathbf{k}_{\perp }, & \uparrow & ,\downarrow
)=+\omega _{1}\omega
_{2}[(q_{1}^{+}+m_{1})(q_{2}^{+}+m_{2})-\mathbf{q}_{\perp }^{2}]/
\sqrt{2}, \\
C_{0}^{F}(x,\mathbf{k}_{\perp }, & \downarrow & ,\uparrow
)=-\omega _{1}\omega
_{2}[(q_{1}^{+}+m_{1})(q_{2}^{+}+m_{2})-\mathbf{q}_{\perp }^{2}]/
\sqrt{2}, \\
C_{0}^{F}(x,\mathbf{k}_{\perp }, & \uparrow & ,\uparrow )=\omega _{1}\omega
_{2}[(q_{1}^{+}+m_{1})q_{2}^{L}-(q_{2}^{+}+m_{2})q_{1}^{L}]/\sqrt{2}, \\
C_{0}^{F}(x,\mathbf{k}_{\perp }, & \downarrow & ,\downarrow
)=\omega _{1}\omega
_{2}[(q_{1}^{+}+m_{1})q_{2}^{R}-(q_{2}^{+}+m_{2})q_{1}^{R}]/\sqrt{
2},
\end{array}
\right.
\end{equation}
which satisfy the relation,
\begin{equation}
\sum_{\lambda _{1},\lambda _{2}}C_{0}^{F}(x,\mathbf{k}_{\perp },\lambda
_{1},\lambda _{2})^{\ast }C_{0}^{F}(x,\mathbf{k}_{\perp },\lambda
_{1},\lambda _{2})=1.  \label{norm}
\end{equation}
We can see that there are also two higher helicity $(\lambda _{1}+\lambda
_{2}=\pm 1)$ components in the expression of the light-cone spin wave
function of the pion besides the ordinary helicity $(\lambda _{1}+\lambda
_{2}=0)$ components. Such higher helicity components \cite%
{Ma93,Huang94,Ma95,Cao97} come from the Melosh-Wigner rotation,
and the same effect plays an important role to understand the
proton \textquotedblleft spin puzzle" in the nucleon case
\cite{Ma91,Ma96}. One may also state that these higher helicity
components contain contribution from orbital angular moment from a
relativistic viewpoint \cite{MS98}. Therefore  we can obtain the
light-cone representation for the spin structure of the pion,
which is the minimal Fock-state of the pion light-cone wave
function:
\begin{equation}
\left\{
\begin{array}{lll}
\Psi _{\pi R}(x,\mathbf{k}_{\perp },\uparrow ,\downarrow
)=+\frac{m}{\sqrt{
2(m^{2}+\mathbf{k}_{\perp }^{2})}}\varphi _{\pi }, & \left[ l^{z}=0\right] &  \\
\Psi _{\pi R}(x,\mathbf{k}_{\perp },\downarrow ,\uparrow
)=-\frac{m}{\sqrt{
2(m^{2}+\mathbf{k}_{\perp }^{2})}}\varphi _{\pi }, & \left[ l^{z}=0\right] &  \\
\Psi _{\pi R}(x,\mathbf{k}_{\perp },\uparrow ,\uparrow
)=-\frac{k_{1}-ik_{2} }{\sqrt{2(m^{2}+\mathbf{k}_{\perp
}^{2})}}\varphi _{\pi }, & \left[ l^{z}=-1\right] &
\\
\Psi _{\pi R}(x,\mathbf{k}_{\perp },\downarrow ,\downarrow
)=-\frac{ k_{1}+ik_{2}}{\sqrt{2(m^{2}+\mathbf{k}_{\perp
}^{2})}}\varphi _{\pi }, & \left[ l^{z}=+1\right] &
\end{array}%
\right.  \label{LCpionWF}
\end{equation}
in which we may employ the Brodsky-Huang-Lepage (BHL) prescription \cite%
{BHL81},
\begin{equation}
\varphi _{\pi }(x,\mathbf{k})=A\exp \left[-\frac{1}{8{\beta
}^{2}}\frac{\mathbf{k}_{\perp }^{2}+m^{2}}{x(1-x)}\right],
\end{equation}%
for the momentum space wave function. Each configuration satisfies
the spin sum rule: $J^{z}=S_{q}^{z}+S_{\overline{q}}^{z}+l^{z}=0$.
Hence, the Fock expansion of the two particle Fock-state for the
pion has four possible spin combinations:
\begin{equation}
\begin{array}{lll}
\left\vert \Psi _{\pi }\left( P^{+},\mathbf{P}_{\perp
}=\mathbf{0}_{\perp }\right) \right\rangle &=&\int
\frac{\mathrm{d}^{2}\mathbf{k}_{\perp }\mathrm{d}x}{16{
\pi }^{3}}  \\
&&\times \left[ \Psi _{\pi R}(x,\mathbf{k}_{\perp },\uparrow
,\downarrow )\left\vert xP^{+},\mathbf{k}_{\perp },\uparrow
,\downarrow \right\rangle +\Psi _{\pi R}(x,\mathbf{k}_{\perp
},\downarrow ,\uparrow )\left\vert xP^{+},
\mathbf{k}_{\perp },\downarrow ,\uparrow \right\rangle \right.    \\
&&\left. +\Psi _{\pi R}(x,\mathbf{k}_{\perp },\uparrow ,\uparrow
)\left\vert xP^{+},\mathbf{k}_{\perp },\uparrow ,\uparrow
\right\rangle +\Psi _{\pi R}(x, \mathbf{k}_{\perp },\downarrow
,\downarrow )\left\vert xP^{+},\mathbf{k}
_{\perp },\downarrow ,\downarrow \right\rangle \right] .   \\
&&
\end{array}
\label{wavefunction}
\end{equation}

To get the Fock-state spin wave function of the pion, we adopt
here another simple way in the full relativistic field theory
treatment of the interaction vertex along with the idea in
\cite{BD80,Bro2001}. We assume that one can consider the pion
vertex connecting to two spin-$\frac{1}{2}$ fermions (e.g. two
quarks) by only taking into account the minimal Fock-state. In the
standard light-cone frame for the pion form factor,
\begin{equation}
\left\{
\begin{array}{lll}
p_{1} & = & (xP^{+},\frac{\mathbf{p}_{1\perp }^{2}+m^{2}}{xP^{+}},
\mathbf{p}_{1\perp }) \\
p_{2} & = & ((1-x)P^{+},\frac{\mathbf{p}_{2\perp }^{2}+m^{2}}{
(1-x)P^{+}},\mathbf{p}_{2\perp }), \\
P & = & (P^{+},\frac{M^{2}}{P^{+}},\mathbf{0}_{\perp }), \\
P^{\prime } & = & (P^{\prime +},\frac{q^{2}+\mathbf{q}_{\perp
}^{2}}{P^{\prime +}},
\mathbf{p^{\prime }}_{\perp }), \\
q & = & (0,\frac{Q^{2}}{P^{+}},\mathbf{q}_{\perp }),
\end{array}
\right.
\end{equation}
we can obtain the above four components of the spin wave function
by calculating the matrix elements of
\begin{equation}
\frac{\overline{u}(p_{1}^{+},p_{1}^{-},\mathbf{k}_{\perp
})}{\sqrt{ p_{1}^{+}}}\gamma
_{5}\frac{v(p_{2}^{+},p_{2}^{-},-\mathbf{k}_{\perp
})}{\sqrt{p_{2}^{+}}},
\end{equation}
from which we have
\begin{equation}
\left\{
\begin{array}{lll}
\frac{\overline{u}_{\uparrow }}{\sqrt{p_{1}^{+}}}\gamma _{5}\frac{
v_{\downarrow }}{\sqrt{p_{2}^{+}}} & = &
+\frac{2mP^{+}}{4mx(1-x)P^{+2}},
 \\
\frac{\overline{u}_{\downarrow }}{\sqrt{p_{1}^{+}}}\gamma
_{5}\frac{ v_{\uparrow }}{\sqrt{p_{2}^{+}}} & = &
-\frac{2mP^{+}}{4mx(1-x)P^{+2}},
  \\
\frac{\overline{u}_{\uparrow }}{\sqrt{p_{1}^{+}}}\gamma _{5}\frac{
v_{\uparrow }}{\sqrt{p_{2}^{\prime +}}} & = & -\frac{2(k_{1}-ik_{2})P^{+}}{
4mx(1-x)P^{+2}},  \\
\frac{\overline{u}_{\downarrow }}{\sqrt{p_{1}^{+}}}\gamma
_{5}\frac{ v_{\downarrow }}{\sqrt{p_{2}^{+}}} & = &
-\frac{2(k_{1}+ik_{2})P^{+}}{ 4mx(1-x)P^{+2}}.
\end{array}
\right.
\end{equation}
After the normalization, we can get the same result of Eq.~(\ref{LCpionWF})
too.

Furthermore, we can get $\left\langle \Psi _{\pi }\left(
P^{+},\mathbf{P}_{\perp }=\mathbf{0}_{\perp }\right) \right\vert $
from the matrix elements of
\begin{equation}
\frac{ \overline{v}(p_{2}^{+},p_{2}^{-},-\mathbf{k}_{\perp
})}{\sqrt{ p_{2}^{+}}}\gamma
_{5}\frac{u(p_{1}^{+},p_{1}^{-},\mathbf{k}_{\perp
})}{\sqrt{p_{1}^{+}}},
\end{equation}
so that
\begin{equation}
\begin{array}{lll}
\left\langle \Psi _{\pi }\left( P^{+},\mathbf{P}_{\perp }\right)
\right\vert & = &
\int \frac{\mathrm{d}^{2}\mathbf{k}_{\perp }\mathrm{d}x}{16{\pi }^{3}} \\
&  & \times \left[ \Psi _{\pi L}(x,\mathbf{k}_{\perp },\uparrow ,\downarrow
)\left\langle xP^{+},\mathbf{k}_{\perp },\uparrow ,\downarrow \right\vert
+\Psi _{\pi L}(x,\mathbf{k}_{\perp },\downarrow ,\uparrow )\left\langle
xP^{+},\mathbf{k}_{\perp },\downarrow ,\uparrow \right\vert \right. \\
&  & \left. +\Psi _{\pi L}(x,\mathbf{k}_{\perp },\uparrow ,\uparrow
)\left\langle xP^{+},\mathbf{k}_{\perp },\uparrow ,\uparrow \right\vert
+\Psi _{\pi L}(x,\mathbf{k}_{\perp },\downarrow ,\downarrow )\left\langle
xP^{+},\mathbf{k}_{\perp },\downarrow ,\downarrow \right\vert \right] , \\
&  &
\end{array}
\end{equation}
%
in which,
\begin{equation}
\left\{
\begin{array}{l}
\Psi _{\pi L}(x,\mathbf{k}_{\perp },\uparrow ,\downarrow )=-\frac{m}{\sqrt{%
2(m^{2}+\mathbf{k}_{\perp }^{2})}}\varphi _{\pi }^{\ast }, \\
\Psi _{\pi L}(x,\mathbf{k}_{\perp },\downarrow ,\uparrow )=+\frac{m}{\sqrt{%
2(m^{2}+\mathbf{k}_{\perp }^{2})}}\varphi _{\pi }^{\ast }, \\
\Psi _{\pi L}(x,\mathbf{k}_{\perp },\uparrow ,\uparrow )=+\frac{k_{1}+ik_{2}%
}{\sqrt{2(m^{2}+\mathbf{k}_{\perp }^{2})}}\varphi _{\pi }^{\ast }, \\
\Psi _{\pi L}(x,\mathbf{k}_{\perp },\downarrow ,\downarrow )=+\frac{%
k_{1}-ik_{2}}{\sqrt{2(m^{2}+\mathbf{k}_{\perp }^{2})}}\varphi _{\pi }^{\ast }.%
\end{array}%
\right.
\end{equation}

Therefore, we can get the pion elastic charge form factor through the
definition: $\left\langle \Psi _{\pi }\left( P^{\prime }\right) \right\vert
J^{+}\left\vert \Psi _{\pi }\left( P\right) \right\rangle \delta ^{3}(%
\mathbf{p}+\mathbf{q}-\mathbf{p^{\prime }})=F_{\pi
}(Q^{2})(P+P^{\prime })$, where $J^{\mu }=\overline{q}e\gamma
^{\mu }e_{q}q$ is the vector current,
\begin{equation}
F_{\pi
^{+}}(Q^{2})=(e_{u}+e_{\overline{d}})\int_{0}^{1}\mathrm{d}x\int
\frac{\mathrm{d}^{2}\mathbf{k}_{\perp }}{16{\pi
}^{3}}\frac{m^{2}+\mathbf{k}_{\perp }\cdot \mathbf{k}_{\perp
}^{\prime }}{\sqrt{m^{2}+\mathbf{k}_{\perp
}^{2}}\sqrt{m^{2}+\mathbf{k}_{\perp }^{\prime 2}}}\varphi _{\pi
}^{\ast }(x,\mathbf{k}_{\perp }^{\prime })\varphi _{\pi
}(x,\mathbf{k}_{\perp }),
\end{equation}%
in which $\mathbf{k^{\prime }}_{\perp }=\mathbf{k}_{\perp
}+(1-x)\mathbf{q}_{\perp }$ for the final state light-cone wave
function when taking into account the Drell-Yan-West assignment
\cite{DYW}.

\section{THE TWO PARTICLE FOCK-STATE OF THE PHOTON VERTEX}

Similar with the pion vertex and with the same assumption, we can also
obtain the spin wave function of the spin-1 photon from the minimal
Fock-state basis of the photon vertex Eq.~(\ref{Photon Fock}) by calculating
the matrix elements of
\begin{equation}
\frac{\overline{v}(p_{2}^{+},p_{2}^{-},\mathbf{p}_{2\perp
})}{\sqrt{ p_{2}^{+}}}\gamma \cdot \epsilon ^{\ast
}\frac{u(p_{1}^{+},p_{1}^{-}, \mathbf{p}_{1\perp
})}{\sqrt{p_{1}^{+}}},
\end{equation}%
which are the numerators of the wave functions corresponding to
each constituent spin $S^{z}$ configuration. The two boson
polarization vectors in light-cone gauge are $\epsilon ^{\mu
}=(\epsilon ^{+}=0,\epsilon ^{-}, \mathbf{\epsilon }_{\perp }),$
where $\mathbf{\epsilon } _{\perp \uparrow ,\downarrow }=\mp
\frac{1}{\sqrt{2}}(\widehat{\mathbf{x}}\pm \widehat{\mathbf{y}}).$
To satisfy the Lorentz condition $k_{photon}\cdot \epsilon =0, $
the polarizations have the relation $\epsilon ^{-}=\frac{2\mathbf{
\epsilon }_{\perp }\cdot \mathbf{k}_{\perp }}{k^{+}}$ with
$k_{hoton} $,
\begin{equation}
\left\{
\begin{array}{lll}
\Psi _{L}^{\uparrow }(x,\mathbf{k}_{\perp },\uparrow ,\downarrow
)=-\frac{ \sqrt{2}(k_{1}-ik_{2})}{1-x}\varphi _{\gamma }, & \left[
l^{z}=+1\right]  &
\\
\Psi _{L}^{\uparrow }(x,\mathbf{k}_{\perp },\downarrow ,\uparrow
)=+\frac{ \sqrt{2}(k_{1}-ik_{2})}{x}\varphi _{\gamma }, & \left[
l^{z}=+1\right]  &
\\
\Psi _{L}^{\uparrow }(x,\mathbf{k}_{\perp },\uparrow ,\uparrow
)=-\frac{
\sqrt{2}m}{x(1-x)}\varphi _{\gamma }, & \left[ l^{z}=0\right]  &  \\
\Psi _{L}^{\uparrow }(x,\mathbf{k}_{\perp },\downarrow ,\downarrow )=0, &  &
\end{array}%
\right.
\end{equation}%
in which:
\begin{equation}
\varphi _{\gamma }=\frac{e_{q}}{D}=\frac{e_{q}}{\lambda
^{2}-\frac{m^{2}+ \mathbf{k}_{\perp
}^{2}}{x}-\frac{m^{2}+\mathbf{k}_{\perp }^{2}}{1-x}},
\end{equation}%
where $\lambda $ is the photon mass and equals to 0. Moreover, as Lepage and
Brodsky had mentioned in \cite{Lep80}, the transition form factor of the
process $\gamma ^{\ast }\gamma \rightarrow \pi ^{0}$ has the energy
denominator
\begin{equation}
D=\lambda ^{2}-\frac{m^{2}+\mathbf{k}_{\perp }^{\prime
2}}{x}-\frac{m^{2}+ \mathbf{k}_{\perp }^{\prime 2}}{1-x},
\end{equation}%
where $\mathbf{k}_{\perp }^{\prime }=\mathbf{k}_{\perp
}+(1-x)\mathbf{q}_{\perp }$ for the final state light-cone wave
function when taking into account the Drell-Yan-West assignment
\cite{DYW}. Each configuration satisfies the spin sum rule:
$J^{z}=S_{q}^{z}+S_{\overline{q}}^{z}+l^{z}=+1.$  Therefore, the
two particle Fock-state for the photon $(J^{z}=+1)$ has four
possible spin combinations:

\begin{eqnarray}
\left\langle \Psi _{\gamma }^{\uparrow }\left( P^{\prime
+},\mathbf{P}_{\perp }^{\prime }\right) \right\vert &=&\int
\frac{\mathrm{d}^{2}\mathbf{k}_{\perp
}\mathrm{d}x}{16{\pi }^{3}}  \nonumber \\
&&\times \left[ \Psi _{L}^{\uparrow }(x,\mathbf{k}_{\perp },\uparrow
,\downarrow )\left\langle xP^{\prime +},\mathbf{k}_{\perp },\uparrow
,\downarrow \right\vert +\Psi _{L}^{\uparrow }(x,\mathbf{k}_{\perp
},\downarrow ,\uparrow )\left\langle xP^{\prime +},\mathbf{k}_{\perp
},\downarrow ,\uparrow \right\vert \right.  \nonumber \\
&&\left. +\Psi _{L}^{^{\uparrow }}(x,\mathbf{k}_{\perp },\uparrow ,\uparrow
)\left\langle xP^{\prime +},\mathbf{k}_{\perp },\uparrow ,\uparrow
\right\vert +\Psi _{L}^{^{\uparrow }}(x,\mathbf{k}_{\perp },\downarrow
,\downarrow )\left\langle xP^{\prime +},\mathbf{k}_{\perp },\downarrow
,\downarrow \right\vert \right] ,  \nonumber \\
\end{eqnarray}%
and respectively we can get the wave function of the photon which
$J^{z}=-1$,
\begin{equation}
\left\{
\begin{array}{lll}
\Psi _{L}^{\downarrow }(x,\mathbf{k}_{\perp },\uparrow ,\downarrow
)=-\frac{
\sqrt{2}(k_{1}+ik_{2})}{x}\varphi _{\gamma }, & \left[ l^{z}=-1\right] &  \\
\Psi _{L}^{\downarrow }(x,\mathbf{k}_{\perp },\downarrow ,\uparrow
)=+\frac{ \sqrt{2}(k_{1}+ik_{2})}{1-x}\varphi _{\gamma }, & \left[
l^{z}=-1\right] &
\\
\Psi _{L}^{\downarrow }(x,\mathbf{k}_{\perp },\uparrow ,\uparrow )=0, &  &
\\
\Psi _{L}^{\downarrow }(x,\mathbf{k}_{\perp },\downarrow
,\downarrow )=- \frac{\sqrt{2}m}{x(1-x)}\varphi _{\gamma }, &
\left[ l^{z}=0\right] &
\end{array}%
\right.
\end{equation}

\begin{eqnarray}
\left\langle \Psi _{\gamma }^{\downarrow }\left( P^{\prime
+},\mathbf{P}_{\perp }^{\prime }\right) \right\vert &=&\int
\frac{\mathrm{d}^{2}\mathbf{k}_{\perp
}\mathrm{d}x}{16{\pi }^{3}}  \nonumber \\
&&\times \left[ \Psi _{L}^{\downarrow }(x,\mathbf{k}_{\perp },\uparrow
,\downarrow )\left\langle xP^{\prime +},\mathbf{k}_{\perp },\uparrow
,\downarrow \right\vert +\Psi _{L}^{\downarrow }(x,\mathbf{k}_{\perp
},\downarrow ,\uparrow )\left\langle xP^{\prime +},\mathbf{k}_{\perp
},\downarrow ,\uparrow \right\vert \right.  \nonumber \\
&&\left. +\Psi _{L}^{\downarrow }(x,\mathbf{k}_{\perp },\uparrow ,\uparrow
)\left\langle xP^{\prime +},\mathbf{k}_{\perp },\uparrow ,\uparrow
\right\vert +\Psi _{L}^{^{\downarrow }}(x,\mathbf{k}_{\perp },\downarrow
,\downarrow )\left\langle xP^{\prime +},\mathbf{k}_{\perp },\downarrow
,\downarrow \right\vert \right] .  \nonumber \\
&&
\end{eqnarray}
By calculating the matrix elements of
\begin{equation}
\frac{\overline{u}(p_{1}^{+},p_{1}^{-},\mathbf{p}_{1\perp
})}{\sqrt{ p_{1}^{+}}}\gamma \cdot \epsilon
\frac{v(p_{2}^{+},p_{2}^{-},\mathbf{p}_{2\perp
})}{\sqrt{p_{2}^{+}}},
\end{equation}
we obtain $\left\vert \Psi _{\gamma }^{\uparrow }\left( P^{\prime
+},\mathbf{P}_{\perp }^{\prime }\right) \right\rangle $, which is
the conjugate part of $\left\langle \Psi _{\gamma }^{\uparrow
}\left( P^{\prime +},\mathbf{P}_{\perp }^{\prime }\right)
\right\vert $

\begin{equation}
\left\{
\begin{array}{l}
\Psi _{R}^{\uparrow }(x,\mathbf{k}_{\perp },\uparrow ,\downarrow
)=-\frac{
\sqrt{2}(k_{1}+ik_{2})}{1-x}\varphi _{\gamma }, \\
\Psi _{R}^{\uparrow }(x,\mathbf{k}_{\perp },\downarrow ,\uparrow
)=+\frac{
\sqrt{2}(k_{1}+ik_{2})}{x}\varphi _{\gamma }, \\
\Psi _{R}^{\uparrow }(x,\mathbf{k}_{\perp },\uparrow ,\uparrow
)=-\frac{
\sqrt{2}m}{x(1-x)}\varphi _{\gamma }, \\
\Psi _{R}^{\uparrow }(x,\mathbf{k}_{\perp },\downarrow ,\downarrow
)=0,
\end{array}%
\right.
\end{equation}

\begin{eqnarray}
\left\vert \Psi _{\gamma }^{\uparrow }\left( P^{\prime
+},\mathbf{P}_{\perp }^{\prime }\right) \right\rangle &=&\int
\frac{\mathrm{d}^{2}\mathbf{k}_{\perp }
\mathrm{d}x}{16{\pi }^{3}}  \nonumber \\
&&\times \left[ \Psi _{R}^{\uparrow }(x,\mathbf{k}_{\perp },\uparrow
,\downarrow )\left\vert xP^{\prime +},\mathbf{k}_{\perp },\uparrow
,\downarrow \right\rangle +\Psi _{R}^{\uparrow }(x,\mathbf{k}_{\perp
},\downarrow ,\uparrow )\left\vert xP^{\prime +},\mathbf{k}_{\perp
},\downarrow ,\uparrow \right\rangle \right.  \nonumber \\
&&\left. +\Psi _{R}^{^{\uparrow }}(x,\mathbf{k}_{\perp },\uparrow ,\uparrow
)\left\vert xP^{\prime +},\mathbf{k}_{\perp },\uparrow ,\uparrow
\right\rangle +\Psi _{R}^{^{\uparrow }}(x,\mathbf{k}_{\perp },\downarrow
,\downarrow )\left\vert xP^{\prime +},\mathbf{k}_{\perp },\downarrow
,\downarrow \right\rangle \right] .  \nonumber \\
&&
\end{eqnarray}

From above calculations, we can arrive at the following
conclusions: when a composite system is transformed from an
ordinary equal-time frame to a light-cone frame, the spin of each
constituent will undergo a Melosh-Wigner rotation, and these spin
rotations for the constituents are not necessarily the same since
the constituents have different internal motions. Therefore, the
sum of the constituents' spin is not Lorentz invariant. For
example, although the pion has only the $\lambda _{1}+\lambda
_{2}=0$ spin components in the rest frame of itself, it may have
$\lambda _{1}+$ $\lambda _{2}=\pm 1$ spin components in the
light-cone frame (infinite-momentum frame), in which $\lambda
_{1}\ $and $\lambda _{2}$ are the quark and the antiquark
helicities, respectively. We could also obtain the similar results
in the Fock expansion of the photon wave function, in which there
are the $\lambda _{1}+\lambda _{2}=0$ helicity states for a vector
particle in the light-cone frame. The similar conclusion is also
true in pure QED case of the electron as a composite system of two
Fock-state particles \cite{Bro2001}. These general results for the
spin structure of composite systems are distinct from the naive
intuitive expectation that the quark spins sum to the hadron spin,
and support the viewpoint that the proton ``spin puzzle" can be
understood as due to the relativistic motion of quarks inside the
nucleon in the light-cone formalism \cite{Ma91,Ma96}.

\section{PION-PHOTON AND PHOTON-PION TRANSITION FORM FACTORS}

In this part, we study the pion-photon and the photon-pion
transition form factors, and perform our theoretical and numerical
analysis respectively. In addition, we discuss the $\pi ^{0}\to
\gamma \gamma $ form factor and the decay width $\mathit{\Gamma
}(\pi \to \gamma \gamma )$ at $Q^{2}=0$. As a matter of fact, we
can find that $F_{\gamma ^{\ast }\pi \to \gamma }(0)=F_{\gamma
^{\ast }\gamma \to \pi }(0)=F_{\pi \to \gamma \gamma }(0)$ for
these three processes at $Q^{2}=0$. Theoretically and generally
speaking, the transition form factor calculated by
Brodsky-Lepage~\cite{Lep80}, Cao-Huang-Ma~\cite{Cao96}, and
Kroll-Raulfs~\cite{Kro96} should be the pion-photon transition
form factor $\gamma ^{\ast }\pi ^{0}\to \gamma $ in the physical
process $e+\pi ^{0}\to e+\gamma $, as $e\to e$ provides the
virtual photon. This should be physically different from the
photon-pion transition form factor $\gamma ^{\ast }\gamma \to \pi
^{0}$ as can be realized in $e+e\to e+e+\pi ^{0}$ or
$e+A(\mbox{Nucleus})\to e+A+\pi ^{0}$. The $\gamma ^{\ast }\gamma
\to \pi ^{0}$ transition form factor has been measured at
Cornell~\cite{CLEO98} and at DESY~\cite{Beh91} through the
$e^{+}+e^{-}\to e^{+}+e^{-}+\pi ^{0}$ process, while the latter
process $e+A(\mbox{Nucleus})\to e+A+\pi ^{0}$ can be performed by
the HERMES Collaboration or other facilities. By employing these
experimental data, Hwang and Choi-Ji discussed the transition form
factors $\gamma ^{\ast }\gamma \to \pi ^{0}$ and $\gamma ^{\ast
}\gamma ^{\ast }\to \pi ^{0}$ theoretically in \cite{Hwang2001}.

\subsection{$\protect\gamma ^{\ast }\protect\pi ^{0}\to \protect\gamma $
TRANSITION FORM FACTOR}

From the pion and the photon vertexes in form of wave functions
that we have got, we can naturally obtain the $\gamma ^{\ast }\pi
^{0}\to \gamma $ transition form factor from its definition. The
form factor $F_{\gamma ^{\ast }\pi \to \gamma },$ in which a pion
is struck by an off-shell photon and decays into an on-shell
photon, is defined by the $\gamma ^{\ast }\pi \gamma $ vertex,
\begin{equation}
\Gamma _{\mu }=-ie^{2}F_{\gamma ^{\ast }\pi \to \gamma }(Q^{2})\varepsilon
_{\mu \nu \rho \sigma }p_{\pi }^{\nu }\epsilon ^{\rho }q^{\sigma },
\end{equation}%
in which $q$ is the momentum of the off-shell photon, $
-Q^{2}=q^{2}=q^{+}q^{-}-\mathbf{q}_{\perp }^{2}=-\mathbf{q}_{\perp
}^{2},$ and $\epsilon $ is the polarization vector of the on-shell
photon.

For the physical state of $\pi^0$, one should also take into
account the color and flavor degrees of freedom into account
\cite{Lep80,Cao96}
\begin{equation}
\left|\Psi_{\pi^0}\right>=\sum_a
\frac{\delta^a_b}{\sqrt{n_c}}\frac{1}{\sqrt{2}} \left[
\left|u^a\bar{u}^b \right>- \left|d^a\bar{d}^b \right>\right],
\end{equation}
where $a$ and $b$ are color indices, $n_c=3$ is the number of
colors, and now $\left|q^a\bar{q}^b \right>$ contains the full
spin structure in Sec.~II. Therefore, we can get
\begin{equation}
\Gamma ^{+}=\left\langle \Psi _{\gamma }^{\uparrow }\left(
P^{\prime +},\mathbf{P}_{\perp }^{\prime }\right) \right\vert
J^{+}\left\vert \Psi _{\pi }\left( P^{+},\mathbf{P}_{\perp
}\right) \right\rangle \delta ^{3}(\mathbf{p}+
\mathbf{q}-\mathbf{p^{\prime }}).
\end{equation}
As a matter of fact, we can get the same result if we use
$\left\langle \Psi _{\gamma }^{\downarrow }\left( P^{\prime
+},\mathbf{P}_{\perp }^{\prime }\right) \right\vert $. Then we
get:
\begin{eqnarray}
F_{\gamma ^{\ast }\pi \to \gamma }(Q^{2}) &=&\frac{\Gamma ^{+}}{
-ie^{2}(\mathbf{\epsilon}_{\perp }\times \mathbf{q}_{\perp })p_{\pi }^{-}}  \nonumber \\
&=&2\sqrt{3}(e_{u}^{2}-e_{d}^{2})\int_{0}^{1}\mathrm{d}x\int
\frac{\mathrm{d} ^{2}\mathbf{k}_{\perp }}{16{\pi }^{3}}\varphi
_{\pi }(x,\mathbf{k}_{\perp })
\nonumber \\
&& \left\{\frac{m}{x\sqrt{ m^{2}+\mathbf{k}_{\perp }^{2}}} \times
\left[ \frac{1}{\frac{m^{2}+\mathbf{k}_{\perp }^{\prime
2}}{x}+\frac{m^{2}+\mathbf{k}_{\perp }^{\prime
2}}{1-x}}\right]+(1\leftrightarrow 2) \right\} \nonumber \\
&=&4\sqrt{3}(e_{u}^{2}-e_{d}^{2})\int_{0}^{1}\mathrm{d}x\int
\frac{\mathrm{d} ^{2}\mathbf{k}_{\perp }}{16{\pi }^{3}} \varphi
_{\pi }(x,\mathbf{k}_{\perp })
\nonumber \\
&&  \frac{m}{x\sqrt{ m^{2}+\mathbf{k}_{\perp }^{2}}}  \times
\left[ \frac{1}{\frac{m^{2}+\mathbf{k}_{\perp }^{\prime
2}}{x}+\frac{m^{2}+\mathbf{k}_{\perp }^{\prime 2}}{1-x}}\right]
.\label{Fpionphoton}
\end{eqnarray}%
The arguments of the final states of the light-cone wavefunction
are $ \mathbf{k}_{\perp i}^{\prime }=\mathbf{k}_{\perp
i}+\mathbf{q}_{\perp }-x_{i}\mathbf{q}_{\perp }$ for the struck
quark, $\mathbf{k}_{\perp i}^{\prime }=\mathbf{k}_{\perp
i}-x_{i}\mathbf{q}_{\perp }$ for the spectator quark after
considering the Drell-Yan-West assignment \cite{DYW}, and the
virtual photon momentum $q_{\mu }$ is specified with $q^{+}=0$ to
eliminate the Z-graph contributions. Therefore, we can obtain the
internal quark transverse momentum of the struck pion
$\mathbf{k}_{\perp }^{\prime }=\mathbf{k}_{\perp }+ (1-x)
\mathbf{q}_{\perp }$ in the center of mass frame.
For the struck particle of the process, we should employ the
Drell-Yan-West assignment in the derivation of the formula for the
transition form factor. The reason is that if we do not know the
wave function of a moving particle (the struck particle after the
virtual photon vertex), we can adopt the wave function in the rest
frame of this particle after using the Drell-Yan-West assignment.
Hence, the rest frame wave functions for the struck particle both
before and after the virtual photon vertex can be used in the
calculation of the transition form factor, even if the struck
particle is in a rest frame before the virtual photon vertex and
in a moving frame after the virtual photon vertex.

Furthermore, we can obtain the $\gamma ^{\ast }\pi ^{0}\to \gamma ^{\ast }$
transition form factor by substituting a virtual photon $\gamma ^{\ast }$
for the on-shell photon $\gamma $, which means to substitute $-Q^{\prime 2}$
for $\lambda ^{2}$ and gives:

\begin{eqnarray}
F_{\gamma ^{\ast }\pi \to \gamma ^{\ast }}(Q^{2},Q^{\prime 2})
&=&4\sqrt{3} (e_{u}^{2}-e_{d}^{2})\int_{0}^{1}\mathrm{d}x\int
\frac{\mathrm{d}^{2}\mathbf{ k}_{\perp }}{16{\pi }^{3}}\varphi
_{\pi }(x,\mathbf{k}_{\perp })\frac{m}{x\sqrt{
m^{2}+\mathbf{k}_{\perp }^{2}}}  \nonumber \\
&&\times \left[ \frac{1}{Q^{\prime 2}+\frac{m^{2}+\mathbf{k}_{\perp
}^{\prime 2}}{x}+\frac{m^{2}+\mathbf{k}_{\perp }^{\prime 2}}{1-x}}\right] .
\end{eqnarray}
The leading order behavior of $F_{\gamma ^{\ast }\pi \to \gamma ^{\ast
}}(Q^{2})$ can be obtained by taking limits of $Q^{2}\to \infty $ and $%
Q^{\prime 2}\to \infty $, thus we can get:
\begin{equation}
F_{\gamma ^{\ast }\pi \to \gamma ^{\ast }}(Q^{2},Q^{\prime 2})=4\sqrt{3}%
(e_{u}^{2}-e_{d}^{2})\int_{0}^{1}\mathrm{d}x\phi _{\pi }(x)\frac{1}{%
xQ^{\prime 2}+(1-x)Q^{2}},
\end{equation}
which is equivalent to the results given in \cite{Cao96}.

\subsection{$\protect\pi ^{0}\to \protect\gamma \protect\gamma $ FORM FACTOR
AND DECAY WIDTH $\mathit{\Gamma }(\protect\pi \to \protect\gamma \protect%
\gamma )$}

From general consideration, the $\pi ^{0}\to \gamma \gamma $ form
factor can not be calculated in the $q^{+}=0$ frame. Other choice
of $q_{\mu }$ will cause contribution from Z-graphs, and it should
give the same result as that in the $q^{+}=0$ case if all graphs
are taken into account \cite{Saw92}. Therefore some more
complicated diagrams should be included. Fortunately, we find that
$F_{\pi \to \gamma \gamma }(0)=F_{\gamma \gamma ^{\ast } \to \pi
}(0)$ when $Q^{2}=0$. It is well known that $F_{\pi \to \gamma
\gamma }(0)$ is related to the two photon partial decay width of
pion $\mathit{\Gamma }(\pi \to \gamma \gamma )$ by \cite{CLEO98}:
\begin{equation}
|F_{\pi \to \gamma \gamma }(0)|^2=\frac{64\pi \mathit{\Gamma }(\pi \to
\gamma \gamma )}{(4\pi \alpha )^{2}M_{\pi }^{3}},
\end{equation}%
where $\alpha $ is the QED coupling constant, $M_{\pi }$ is the
pion mass and $\mathit{\Gamma }(\pi \to \gamma \gamma )$ is the
two-photon partial width of the pion. In addition, we can obtain:

\begin{eqnarray}
F_{\gamma \gamma ^{\ast }\to \pi }(0) &=&F_{\pi \to \gamma \gamma }(0)=4%
\sqrt{3}(e_{u}^{2}-e_{d}^{2})\int_{0}^{1}\mathrm{d}x\int \frac{\mathrm{d}^{2}%
\mathbf{k}_{\perp }}{16{\pi }^{3}}  \nonumber \\
&&\left[ \varphi _{\pi }(x,\mathbf{k}_{\perp })\frac{m}{x\sqrt{m^{2}+\mathbf{k}_{\perp }^{2}}}%
\times \frac{1}{\frac{m^{2}+\mathbf{k}_{\perp }^{2}}{x}+\frac{m^{2}+\mathbf{k%
}_{\perp }^{2}}{1-x}}\right] .
\end{eqnarray}
Hence, we can get $\mathit{\Gamma }(\pi \to \gamma \gamma )$ from above
calculations.

\subsection{$\protect\gamma ^{\ast }\protect\gamma \to \protect\pi ^{0}$
TRANSITION FORM FACTOR}

By using the Fock expansions of the pion and the photon wave
functions that we have got above, we can naturally obtain the
$\gamma ^{\ast }\gamma \to \pi ^{0}$ transition form factor from
its definition. The form factor $F_{\gamma ^{\ast }\gamma \to \pi
^{0}},$ in which an on-shell photon is struck by an off-shell
photon and decays into a pion, is defined by the $\gamma ^{\ast
}\gamma \pi $ vertex,
\begin{equation}
\Gamma _{\mu }=-ie^{2}F_{\gamma ^{\ast }\gamma \to \pi
^{0}}(Q^{2})\varepsilon _{\mu \nu \rho \sigma }p_{\pi }^{\nu }\epsilon
^{\rho }q^{\sigma },
\end{equation}%
in which $q$ is the momentum of the off-shell photon,
$-Q^{2}=q^{2}=q^{+}q^{-}-\mathbf{q}_{\perp
}^{2}=-\mathbf{q}_{\perp }^{2},$ and $\epsilon $ is the
polarization vector of the on-shell photon. In the light-cone
frame:
\begin{equation}
\left\{
\begin{array}{lll}
p_{1} & = & (xP^{+},\frac{\mathbf{k}_{\perp }^{2}+m^{2}}{xP^{+}},
\mathbf{k}_{\perp }) \\
p_{2} & = & ((1-x)P^{+},\frac{\mathbf{k}_{\perp
}^{2}+m^{2}}{(1-x)P^{+}},-
\mathbf{k}_{\perp }), \\
p_{1}^{\prime } & = & (xP^{\prime +},\frac{\mathbf{k}_{\perp }^{\prime
2}+m^{2}}{xP^{+}},\mathbf{k^{\prime }}_{\perp }), \\
P_{\gamma } & = & (P^{+},\frac{q^{2}+\mathbf{q}_{\perp
}^{2}}{P^{+}},\mathbf{0
}_{\perp }), \\
P_{\pi }^{\prime } & = & (P^{\prime +},\frac{M^{2}}{P^{\prime +}},
\mathbf{q}_{\perp }), \\
q & = & (0,\frac{Q^{2}}{P^{+}},\mathbf{q}_{\perp }),
\end{array}
\right.
\end{equation}
so we can get
\begin{eqnarray}
F_{\gamma \gamma ^{\ast }\to \pi }(Q^{2}) &=&\frac{\Gamma ^{+}}{
-ie^{2}(\mathbf{\epsilon}_{\perp }\times \mathbf{q}_{\perp
})p_{\pi }^{\prime -}} \nonumber
\\
&=&-4\sqrt{3}(e_{u}^{2}-e_{d}^{2})\int_{0}^{1}\mathrm{d}x\int
\frac{\mathrm{d }^{2}\mathbf{k}_{\perp }}{16{\pi }^{3}}\left[
\varphi _{\pi }(x,\mathbf{k}_{\perp }^{\prime
})\frac{m}{x\sqrt{m^{2}+\mathbf{k}_{\perp }^{\prime 2}}}\right.
\nonumber
\\
&&\left. \times \frac{1}{\lambda
^{2}-\frac{m^{2}+\mathbf{k}_{\perp }^{2}}{x}
-\frac{m^{2}+\mathbf{k}_{\perp }^{2}}{1-x}}\right] ,
\label{Fphotonpion}
\end{eqnarray}%
in which $\mathbf{k}_{\perp }^{\prime }=\mathbf{k}_{\perp
}+(1-x)\mathbf{q}_{\perp }$ after considering the Drell-Yan-West
formula, and $\lambda $ $(=0)$ is the photon mass.

Moreover, we can get the equivalent $F_{\gamma \gamma ^{\ast }\pi
}(Q^{2})$ if we choose another different light-cone frame:
\begin{equation}
\left\{
\begin{array}{lll}
p_{1} & = & (xP^{+},\frac{(x\mathbf{P}_{\perp }+\mathbf{k} _{\perp
})^{2}+m^{2}}{xP^{+}},x\mathbf{P}_{\perp }+\mathbf{k}
_{\perp }) \\
p_{2} & = & ((1-x)P^{+},\frac{((1-x)\mathbf{P}_{\perp }-%
\mathbf{k}_{\perp })^{2}+m^{2}}{(1-x)P^{+}},(1-x)\mathbf{P}
_{\perp }-\mathbf{k}_{\perp }), \\
p_{1}^{\prime } & = & (xP^{\prime +},\frac{\mathbf{k}_{\perp }^{\prime
2}+m^{2}}{xP^{+}},\mathbf{k^{\prime }}_{\perp }), \\
P_{\gamma } & = & (P^{+},\frac{q^{2}+\mathbf{q}_{\perp
}^{2}}{P^{+}},\mathbf{P
}_{\perp }), \\
P_{\pi }^{\prime } & = & (P^{\prime +},\frac{M^{2}}{P^{\prime +}},
\mathbf{0}_{\perp }), \\
q & = & (0,\frac{Q^{2}}{P^{+}},\mathbf{q}_{\perp }),
\end{array}
\right.
\end{equation}
in which $\mathbf{P}_{\perp }=-\mathbf{q}_{\perp }$, and $
\mathbf{k}_{\perp }^{\prime }=\mathbf{k}_{\perp
}+(1-x)\mathbf{q}_{\perp }$.
\begin{equation}
F_{\gamma \gamma ^{\ast }\to \pi
}(Q^{2})=-4\sqrt{3}(e_{u}^{2}-e_{d}^{2})
\int_{0}^{1}\mathrm{d}x\int \frac{\mathrm{d}^{2}\mathbf{k}_{\perp
}}{16{\pi } ^{3}}\varphi _{\pi }(x,\mathbf{k}_{\perp }^{\prime
})\frac{m}{x\sqrt{m^{2}+\mathbf{k}_{\perp }^{\prime
2}}}\frac{1}{D},
\end{equation}
where
\begin{eqnarray}
D &=&P_{\perp }^{2}-\frac{(x\mathbf{P}_{\perp }+\mathbf{k} _{\perp
})^{2}+m^{2}}{x}-\frac{\left((1-x)\mathbf{P}_{\perp }-
\mathbf{k}_{\perp }\right)^{2}+m^{2}}{(1-x)}  \nonumber \\
&=&-\frac{\mathbf{k}_{\perp
}^{2}+m^{2}}{x}-\frac{\mathbf{k}_{\perp }^{2}+m^{2}}{(1-x)}.
\end{eqnarray}%
Therefore, it is obvious that $F_{\gamma \gamma ^{\ast } \to \pi
}(Q^{2}) $ we have got is frame invariant. In the first frame, we
employed the Drell-Yan-West formula in the process of obtaining
$F_{\gamma \gamma ^{\ast } \to \pi }(Q^{2})$, while in the other
frame we did not use the Drell-Yan-West formula. We want to
emphasize that the Drell-Yan-West formula is more convenient and
simple although these two methods can give the same results.

Furthermore, we can obtain the $\gamma ^{\ast }\gamma ^{\ast }\to \pi ^{0}$
transition form factor by substituting a virtual photon $\gamma ^{\ast }$
for the on-shell photon $\gamma $, which means to substitute $-Q^{\prime 2}$
for $\lambda ^{2}$ and gives:

\begin{eqnarray}
F_{\gamma ^{\ast }\gamma ^{\ast }\to \pi ^{0}}(Q^{2},Q^{\prime 2})
&=&4\sqrt{ 3}(e_{u}^{2}-e_{d}^{2})\int_{0}^{1}\mathrm{d}x\int
\frac{\mathrm{d}^{2} \mathbf{k}_{\perp }}{16{\pi }^{3}}\varphi
_{\pi }(x,\mathbf{k}_{\perp }^{\prime })
\frac{m}{x\sqrt{m^{2}+\mathbf{k}_{\perp }^{\prime 2}}}  \nonumber \\
&&\times \left( \frac{1}{Q^{\prime
2}+\frac{m^{2}+\mathbf{k}_{\perp }^{2}}{x}
+\frac{m^{2}+\mathbf{k}_{\perp }^{2}}{1-x}}\right) .
\end{eqnarray}

\subsection{NUMERICAL CALCULATIONS}

In the formula of the transition form factor $F_{\gamma \gamma
^{\ast}\to \pi }(Q^{2})$, the parameters are the normalization
constant $A$, the harmonic scale $\beta $, and the quark masses
$m$. In order to take a numerical calculation of the transition
form factor $F_{\gamma \gamma ^{\ast }\to \pi }(Q^{2})$ and
compare it with the available experimental data, we need to employ
three constraints to fix those three parameters above. Thus, we
can determine all these three parameters in the transition form
factor uniquely.

\noindent 1.~The weak decay constant $f_{\pi}=92.4$~MeV defined
\cite{decay} from $\pi \to \mu \nu $ decay, thus one can obtain
Eq.~(\ref{decayA}) and Eq.~(\ref{decayB}) from our former work
\cite{Ma93, Xiao2002} or Hwang's paper \cite{Hwang2001}
respectively:
\begin{equation}
\frac{f_{\pi }}{2\sqrt{3}}=\int_{0}^{1}dx\int
\frac{d^{2}\mathbf{k}_{\perp }}{16{\pi }
^{3}}\frac{(k_{1}^{+}+m)(k_{2}^{+}+m)-{\mathbf{k}_{\perp
}}^{2}}{{[(k_{1}^{+}+m)^{2}+{ \mathbf{k}_{\perp
}}^{2}]}^{1/2}{[(k_{2}^{+}+m)^{2}+{\mathbf{k}_{\perp
}}^{2}]}^{1/2}}{\varphi}_{\pi}(x,\mathbf{k}_{\perp }),
\label{decayA}
\end{equation}%
\begin{equation}
\frac{f_{\pi }}{2\sqrt{3}}=\int_{0}^{1}dx\int
\frac{d^{2}\mathbf{k}_{\perp }}{16{\pi }
^{3}}\frac{m}{\sqrt{{m}^{2}+{{\mathbf{k}_{\perp }}^{2}}}}{\varphi
}_{\pi}(x,\mathbf{k}_{\perp }). \label{decayB}
\end{equation}
One can prove that Eq.~(\ref{decayA}) is identical to
Eq.~(\ref{decayB}) after some deduction, by using $q^{+}_i=x_i
{M}$ and ${M}=(m^2+\mathbf{k}_{\perp }^{2})/x(1-x)$ \cite{Ma93}.

\noindent 2.~The charged mean square radius of $\pi ^{+}$ is defined as:
\begin{equation}
\langle r_{\pi ^{+}}^{2}\rangle =-6\frac{\partial F_{\pi
^{+}}(Q^{2})}{
\partial Q^{2}}|_{Q^{2}=0}.  \label{pi+}
\end{equation}
We can find the experimental value of $\langle r_{\pi
^{+}}^{2}\rangle =0.439\pm 0.003~\mbox{ fm}^{2}.$\cite{data1}.

\noindent 3.~The decay width $\mathit{\Gamma }(\pi ^{0}\to \gamma
\gamma )$ has the following relationship with $F_{\pi \gamma
\gamma }(0)\ $and $ F_{\gamma \gamma ^{\ast }\pi }(0)$
\cite{CLEO98}:

\begin{equation}
\left\vert F_{\gamma \gamma ^{\ast }\to \pi }(0)\right\vert
^{2}=\left\vert F_{\pi \to \gamma \gamma }(0)\right\vert
^{2}=\frac{64\pi \mathit{\Gamma } (\pi ^{0}\to \gamma \gamma
)}{(4\pi \alpha )^{2}M_{\pi }^{3}},
\end{equation}
and we could use $\mathit{\Gamma }(\pi ^{0}\to \gamma \gamma
)=7.74\pm 0.54~ {\mathrm eV}$ \cite{CLEO98}, which leads to
$F_{\pi \gamma \gamma }(0)=0.27\pm 0.01~\mbox{ GeV}^{-1}$ in our
calculation.

Therefore, we can obtain $m=200$~MeV (\textit{e.g.}, for the up
quark or the down quark, assuming $m_{u}=m_{d}=m$), $\beta
=410$~MeV, and $A=0.0475~\mbox{ MeV}^{-1}$. Reversely, we can
compute the values of $f_{\pi }$, $\langle r_{\pi ^{+}}^{2}\rangle
$, and $\mathit{\Gamma }(\pi \to \gamma \gamma )$ by using the
above four parameters:
\begin{eqnarray}
f_{\pi } &=&92.4 ~\mbox{ MeV}, \\
\langle r_{\pi ^{+}}^{2}\rangle &=&0.441~\mbox{ fm}^{2}, \\
F_{\pi \to \gamma \gamma }(0) &=&0.271~\mbox{  GeV}^{-1}, \\
\mathit{\Gamma }(\pi \to \gamma \gamma )&=&7.56~\mbox{ eV}.
\end{eqnarray}

\begin{figure}[tbh]
\includegraphics{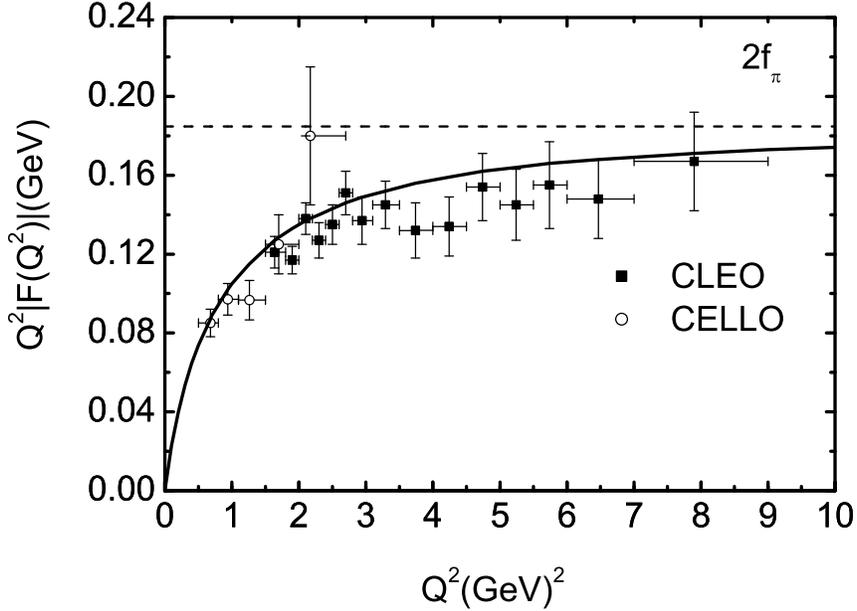}
\caption[*]{\baselineskip13pt Theoretical prediction for the
results of $Q^{2}|F_{\protect\gamma \protect\gamma ^{\ast }\to
\protect\pi }(Q^{2})|$ calculated with the pion wave function in
the BHL prescription compared with the experimental data. The data
for the transition form factor are taken from
Ref.~\protect\cite{CLEO98} and Ref.~\protect\cite{Beh91}.}
\label{transition}
\end{figure}
The results are in good agreement with the experimental data which
we have listed above. Moreover, it is interesting to notice that
the masses of the light-flavor quarks (the up and down quarks)
from the above constrains are just in the correct range
(\textit{e.g.}, $200 \sim 300$~MeV) of the constituent quark
masses from more general considerations. Naturally, the transition
form factor results emerging from this assumption are in quite
good agreement with the experimental data.

Fig.~1 indicates that the theoretical values of the photon-pion
$(\gamma \gamma ^{\ast }\to \pi )$ transition form factors fit the
experimental data well, especially in the case of low $Q^{2}$. One
may consider this work as a light-cone version of relativistic
quark model \cite{Ma93,Xiao2002,LCQM}, which should be valid in
the low-energy scale about $Q^{2}\leq 2~\mbox{ GeV}^{2}$. However,
it is also physically in accordance with the light-cone
perturbative QCD approach \cite{Lep80,Cao96}, which is applicable
at the high-energy scale of $Q^{2}>2~\mbox{ GeV}^{2}$. The reason
is that the hard-gluon exchange between the quark and the
antiquark of the meson, which should be generally considered at
high $Q^{2}$ for exclusive processes, is not necessary to be
incorporated in the light-cone perturbative QCD approach for the
pion-photon transition form factor \cite{Lep80,Cao96}. As a
result, there is no wonder that our predictions for the transition
form factor at high $Q^{2}$ also agree with the experimental data
at high energy scale. Moreover, our numerical result fits the
perturbative QCD result well in the limit $Q^{2}\to \infty $ which
was first given by Brodsky and Lepage \cite{Lep80,Bro81}. Brodsky
and Lepage found that
\begin{equation}
\lim_{Q^{2}\to \infty }Q^{2}|F_{\gamma \gamma ^{\ast }\to \pi
}(Q^{2})|=2f_{\pi },
\end{equation}
in which $f_{\pi }$ is the decay constant. This result predicts
that any mesonic wave function evolves to asymptotic wave function
in the limit $Q^{2}\to \infty $. To describe the soft
nonperturbative region of $Q^{2}$ with a simple interpolation
between $Q^{2}\to \infty $ and $Q^{2}\to 0$ limits, they have
proposed
\begin{equation}
F_{\gamma \gamma ^{\ast }\to \pi }(Q^{2})=\frac{1}{4\pi ^{2}f_{\pi
}}\frac{1}{1+\frac{Q^{2}}{8\pi ^{2}f_{\pi }^{2}}}  \label{soft}
\end{equation}
in \cite{Bro81}. We also discover that our numerical prediction
for $Q^{2}|F_{\gamma \gamma ^{\ast }\to \pi }(Q^{2})|$ is quite
consistent with the numerical result~\cite{CLEO98} of
Eq.~(\ref{soft}), see e.g., Fig.~21 of \cite{CLEO98}.

\begin{figure}[tbh]
\includegraphics{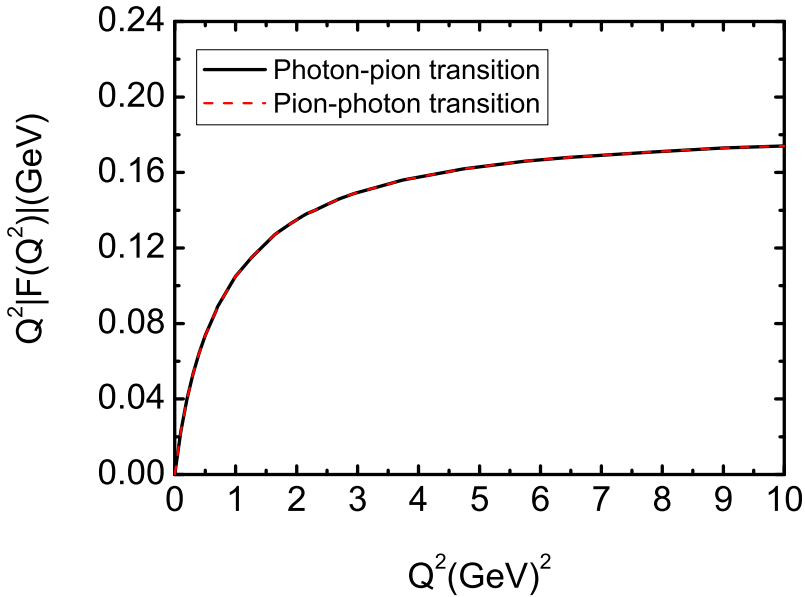}
\caption[*]{\baselineskip13pt Numerical results of $\protect\gamma
\protect \gamma ^{\ast }\to \protect\pi $ transition form factor
compared with the $\protect\gamma ^{\ast }\protect\pi \to
\protect\gamma $ transition form factor, represented by the dashed
and solid lines respectively.} \label{transition2}
\end{figure}

To illustrate the intrinsic relation between the $\gamma \gamma
^{\ast }\to \pi $ and the $\gamma ^{\ast }\pi \to \gamma $
transition form factors, we also carry out the numerical
calculation of $F_{\gamma ^{\ast }\pi \to \gamma }(Q^{2})$
compared with the numerical results of $F_{\gamma \gamma ^{\ast
}\to \pi }(Q^{2})$, as shown in Fig.~2. We find an interesting
result that $F_{\gamma ^{\ast }\pi \to \gamma }(Q^{2})$ is equal
to$\ F_{\gamma \gamma ^{\ast }\to \pi }(Q^{2})$ when $Q^{2}$
varies from $0~\mbox{ GeV}^{2}$ to $10~\mbox{ GeV}^{2}$. Although
$F_{\gamma \gamma ^{\ast }\to \pi }(Q^{2})$ and $F_{\gamma ^{\ast
}\pi \to \gamma }(Q^{2})$ are physically different in the formulae
as we have shown above, Fig.~2 indicates that they are numerically
identical at very high precision, which imply that $F_{\gamma
\gamma ^{\ast } \to \pi }(Q^{2})$ and $F_{\gamma ^{\ast }\pi \to
\gamma }(Q^{2})$ have the same $Q^{2}$ dependence.  The
identification of Eq.~(\ref{Fpionphoton}) and
Eq.~(\ref{Fphotonpion}) can be proved by making the variable
transformation: $\mathbf{k}_{\perp } \to \mathbf{k}_{\perp }-(1-x)
\mathbf{q}_{\perp }$ and then $\mathbf{q}_{\perp } \to
-\mathbf{q}_{\perp }$ for Eq.~(\ref{Fphotonpion}). In fact, the
identification of the two form factors can be understood from
time-reverse invariance as well as parity symmetry for QED and QCD
processes. Therefore, we can safely arrive at another important
conclusion that $F_{\gamma ^{\ast }\gamma ^{\ast }\to \pi
}(Q^{2},Q^{\prime 2})$ and $F_{\gamma ^{\ast }\pi \to \gamma
^{\ast }}(Q^{2},Q^{\prime 2})$ should be numerically identical as
well as having the same $Q^{\prime 2}$ dependence, unless novel
new physics beyond QED and QCD will be involved. This conclusion
is consistent with the results in \cite{Cao96} (for $F_{\gamma
^{\ast }\pi \to \gamma ^{\ast }}(Q^{2},Q^{\prime 2})$) and
\cite{Hwang2001} (for $F_{\gamma ^{\ast }\gamma ^{\ast }\to \pi
}(Q^{2},Q^{\prime 2})$) when taking the $Q^{2}\to \infty $ and
$Q^{\prime 2}\to \infty $ limits,
\begin{equation}
F_{\gamma ^{\ast }\pi \to \gamma ^{\ast }}(Q^{2},Q^{\prime
2})=F_{\gamma ^{\ast }\gamma ^{\ast }\to \pi }(Q^{2},Q^{\prime
2})=4\sqrt{3} (e_{u}^{2}-e_{d}^{2})\int_{0}^{1}\mathrm{d}x\phi
_{\pi }(x)\frac{1}{ xQ^{\prime 2}+(1-x)Q^{2}}.
\end{equation}

\section{CONCLUSIONS}

The light-cone formalism provides a convenient framework for the
relativistic description of hadrons in terms of quark and gluon
degrees of freedom, and for the application of perturbative QCD to
exclusive processes. In the application of the light-cone quark
model, we obtain the minimal Fock-state expansions of the pion and
the photon wave functions from the light-cone representation of
the spin structure of pion and photon vertexes, then we
investigate the pion-photon and the photon-pion transition form
factors of the processes $\gamma ^{\ast }\pi ^{0}\to \gamma $ and
$\gamma ^{\ast }\gamma \to \pi ^{0}$ by employing the
quark-antiquark wave functions of the pion and the photon that we
have obtained. We employ the experimental values of the pion decay
constant $f_{\pi }$, the electromagnetic charged mean squared
radius $\langle r_{\pi ^{+}}^{2}\rangle $, and the decay width
$\mathit{\Gamma }(\pi ^{0}\to \gamma \gamma )$ as the constraints
to fix those three parameters in the pion wave function. With the
fixed pion wave function, we find that our numerical prediction
for the $\gamma ^{\ast }\gamma \to \pi ^{0}$ transition form
factor agrees with the experimental data at low and moderately
high energy scale. Furthermore, we make the numerical comparison
between the transition form factors of $\gamma ^{\ast }\pi ^{0}\to
\gamma $ and $\gamma ^{\ast }\gamma \to \pi ^{0}$, which gives the
result that these two transition form factors are equal and
indicates that these two processes have intrinsic relation while
they are physically different. In addition, we also give the
formulae of $F_{\gamma ^{\ast }\gamma ^{\ast }\to \pi
}(Q^{2},Q^{\prime 2})$ and $F_{\gamma ^{\ast }\pi \to \gamma
^{\ast }}(Q^{2},Q^{\prime 2})$, and find that they should have the
same $Q^{2}$ and $Q^{\prime 2}$ dependence, unless novel new
physics beyond QED and QCD will be involved.

\textbf{ACKNOWLEDGMENTS} This work is partially supported by National
Natural Science Foundation of China under Grant Numbers 10025523 and
90103007. It is also supported by Hui-Chun Chin and Tsung-Dao Lee Chinese
Undergraduate Research Endowment (CURE) at Peking University.

\appendix\textbf{APPENDIX: DRELL-YAN-WEST ASSIGNMENT FOR SPIN PART}

In this appendix, we prove the applicability of the Drell-Yan-West
assignment to the spin part as well as for the denominator.
Supposing $P$ is the momentum of the incident particle at rest,
$P^{\prime }$ is the momentum of the final particle, $p_{1}$ and
$p_{2}$ are for quarks, then $p_{1}$ is struck by $q$ into
$p_{1}^{\prime }$, so we have
\begin{equation}
\left\{
\begin{array}{lll}
p_{1}^{\prime } & = & p_{1}+q, \\
p_{2}^{\prime } & = & p_{2}, \\
P^{\prime } & = & P+q.
\end{array}%
\right.
\end{equation}
The Drell-Yan-West assignment is to change the kinematics for the
final particle from a moving frame to its rest frame:
\begin{equation}
\left\{
\begin{array}{lll}
p_{1}^{\prime } & \to & p_{1}+q -xq= p_{1}+ (1-x) q, \\
p_{2}^{\prime } & \to & p_{2}-(1-x)q, \\
P^{\prime } & \to & P.
\end{array}
\right.
\end{equation}
For the denominators, the general time-ordered field theory framework and
the Drell-Yan-West assignment give the same results respectively:
\begin{eqnarray}
D &=&\lambda ^{2}+\mathbf{q}_{\perp
}^{2}-\frac{m^{2}+(\mathbf{k}_{\perp }+\mathbf{q}_{\perp
})^{2}}{x}-
\frac{m^{2}+\mathbf{k}_{\perp }^{2}}{1-x}, \\
D &=&\lambda ^{2}-\frac{m^{2}+\mathbf{k}_{\perp }^{\prime
2}}{x}-\frac{m^{2}+\mathbf{k}_{\perp }^{\prime 2}}{1-x},
\end{eqnarray}%
where $\mathbf{k}_{\perp }^{\prime }=\mathbf{k}_{\perp
}+(1-x)\mathbf{q}_{\perp }$.

For the spin part, we can prove that the general field theory
framework and the Drell-Yan-West assignment present the identical
formulae, and the assignment is more straightforward to obtain the
results. Taking the calculation of $\left\vert \Psi _{\pi }\left(
P^{+},\mathbf{P}_{\perp }\right) \right\rangle $ for example in
the general framework:

\begin{equation}
\frac{\overline{u}(p_{1}^{+},p_{1}^{-},\mathbf{k}_{\perp }+
\mathbf{q}_{\perp })}{\sqrt{p_{1}^{+}}}\gamma _{5}\frac{
v(p_{2}^{+},p_{2}^{-},-\mathbf{k}_{\perp })}{\sqrt{p_{2}^{+}}},
\end{equation}
we obtain
\begin{equation}
\left\{
\begin{array}{lll}
\frac{\overline{u}_{\uparrow }}{\sqrt{p_{1}^{+}}}\gamma _{5}\frac{
v_{\downarrow }}{\sqrt{p_{2}^{+}}} & = & +\frac{2mP^{+}}{4mx(1-x)P^{+2}}, \\
\frac{\overline{u}_{\downarrow }}{\sqrt{p_{1}^{+}}}\gamma
_{5}\frac{
v_{\uparrow }}{\sqrt{p_{2}^{+}}} & = & -\frac{2mP^{+}}{4mx(1-x)P^{+2}}, \\
\frac{\overline{u}_{\uparrow }}{\sqrt{p_{1}^{+}}}\gamma _{5}\frac{
v_{\uparrow }}{\sqrt{p_{2}^{\prime +}}} & = &
-\frac{2\left(k_{L}+(1-x)q_{L}\right)P^{+}
}{4mx(1-x)P^{+2}}, \\
\frac{\overline{u}_{\downarrow }}{\sqrt{p_{1}^{+}}}\gamma
_{5}\frac{ v_{\downarrow }}{\sqrt{p_{2}^{+}}} & = &
-\frac{2\left(k_{R}+(1-x)q_{R}\right)P^{+}}{ 4mx(1-x)P^{+2}},
\end{array}
\right.
\end{equation}
which is equivalent to the Drell-Yan-West assignment of making the
substitution of $\mathbf{k}_{\perp }^{\prime }=\mathbf{k}_{\perp
}+(1-x)\mathbf{q}_{\perp }$ for $\mathbf{k}_{\perp }$.


\end{document}